\documentclass{elsarticle}

\usepackage{hyperref}

\begin{document}

\begin{frontmatter}

\title{Optimized sample addressing in prism-coupled surface plasmon resonance experiments}

\author{Josu Martinez-Perdiguero\fnref{myfootnote}}
\fntext[myfootnote]{Email address: jesus.martinez@ehu.eus}

\address{Department of Condensed Matter Physics, Faculty of Science and Technology, University of the Basque Country UPV/EHU, Barrio Sarriena s/n, 48940 Leioa, Spain}

\author{Ibon Alonso}
\address{Department of Applied Physics II, Faculty of Science and Technology, University of the Basque Country UPV/EHU, Barrio Sarriena s/n, 48940 Leioa, Spain}

\begin{abstract}
In this work we study the walk-off of the beam from the interrogation spot during rotation in surface plasmon resonance experiments using prism-based coupling such as the widespread Kretschmann configuration. The impossibility of maintaining a stationary footprint on the sensing surface with a fixed rotation axis can be of high importance. This would be specially so if samples are not homogeneous such as in arrays for multiplexing. By theoretically analyzing the behavior of the walk-off during rotation around an arbitrary fixed axis, we find an optimal and simple configuration to minimize this effect. The proposed setup is experimentally tested to verify the results and to show its ease of implementation. Interestingly, the conclusions reached  may also be applied to other techniques employing reflection prisms.
\end{abstract}


\end{frontmatter}


\section{Introduction}
The most standard method for the generation of surface plasmon polaritons for sensing is the prism-coupling of light to a metallic thin-layer using the Kretschmann configuration \cite{kretschmann_e._radiative_1968}. Other methods are also well established, such as the also prism-based Otto configuration \cite{otto_excitation_1968}, gratings \cite{raether_surface_1988} or optical waveguides \cite{dostalek_surface_2001}. However, virtually all commercial systems make use of the former because of its ease of integration, robustness and sensitivity \cite{schasfoort_handbook_2017}.

Based on the Kretschmann configuration, the so-called sensorgram employed to represent surface changes vs. time can be obtained by logging the SPR angle dip, the reflectivity change or the wavelength shift \cite{schasfoort_handbook_2017}. Moreover, a SPR system can work in a fixed angle setup, angle or wavelength scanning setups or use a fan-shaped beam \cite{schasfoort_handbook_2017}. Also, by means of a extense beam and CCD camera, the widespread `imaging' SPR (iSPR) systems have made appearance allowing to monitor large areas and boosting the array capabilities of SPR sensors \cite{campbell_spr_2007}.

\begin{figure}
\begin{center}
  \includegraphics[width=\linewidth]{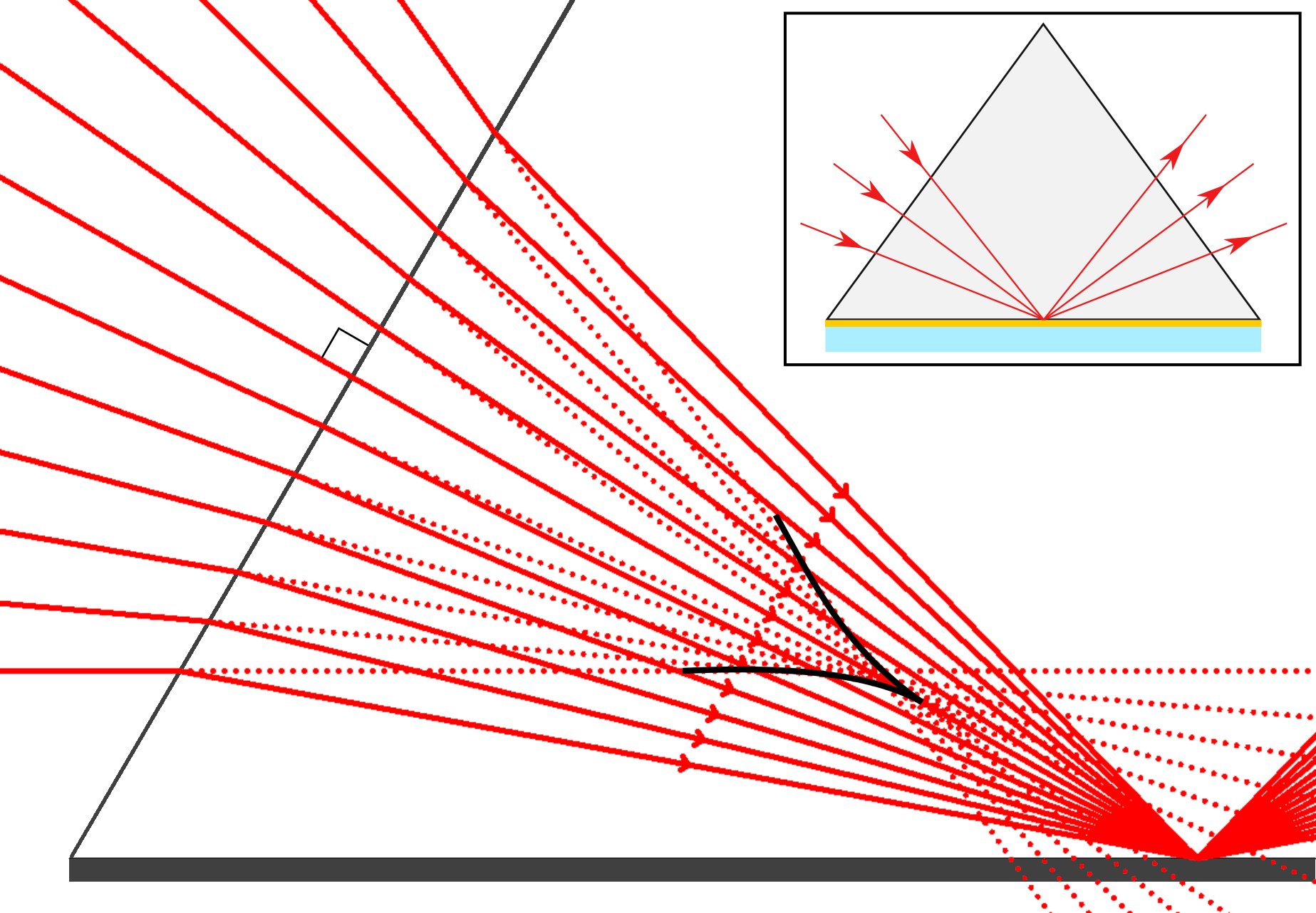}
\end{center}
\caption{Rays arriving at a given point on the base of a prism after refraction on the side face. The extensions of the incoming rays (dotted lines) give rise to a caustic (dark curve). Inset: oversimplified and inaccurate scheme quite commonly employed in the literature on SPR to illustrate the path of light as the prism is rotated.}
\label{fig:caustic}
\end{figure}

In any of the above-mentioned possibilities, using a most common triangular or dove prism, light must cross a face of the prism, be totally reflected at its base (typically coated with a gold thin-film plus adhesion layer), and travel out through the other face towards the detector. At first thought, this seems like a straightforward process and it is commonly oversimplified (even in specialized journals) and wrongly depicted as in the inset of Fig. \ref{fig:caustic}. The most striking error of omission is the missing refraction at the lateral prism faces. This complication is sometimes hidden by only depicting a ray at normal incidence. However, this simple picture of the SPR setup is misleading and the real optical configuration and alignment are more complicated, even more so in setups employed for imaging.

It is customary for SPR systems to implement a $\theta:2\theta$ rotation stage for angle-scanning or angle adjustment. In this case, the rotation of the prism around an axis leads to a change in the optical path of the rays which, if not thoroughly corrected, results in a shift in the beam position at the base of the prism, with the consequent walk-off of the interrogation spot in the sample, and an increase of the area interrogated as the spot deforms along the plane of incidence.

The walk-off issue has been addressed in the literature by a few authors \cite{ulrich_measurement_1973,horn_plasmon_2010,qi_optimized_2012,fontana_maintaining_2013,singh_fixed_2014}, who also proposed configurations to correct it. However, the solutions are either not optimal and/or require high-precision repositioning of optical elements for each angle of incidence resulting in time-consuming and difficult-to-implement experimental setups.

The walk-off of the beam over the sensing surface  during angular interrogation is due to the different optical paths followed by rays at different angles of incidence stemming from the refraction at the air-glass interface. This walk-off can lead to inaccuracies when the sensing surface is non-homogeneous. For example, if the sensing surface is covered by a film, defects and thickness inhomogeneities can result in difficult to detect measurement errors. More importantly, during the last years, the implementation of a high density of differently functionalized spots in the sample surface is being carried out to take advantage of the arraying capabilities of SPR systems, i.e., multiplexing (see, for example, \cite{piliarik_new_2005,ouellet_parallel_2010,lakayan_design_2019}). In this case, an uncontrolled walk-off can result in the interrogation of non-functionalized sensor surface or even of another spot of the array.

In this paper we present a detailed and complete analytical study of the interrogation spot walk-off in prism-based SPR systems during angular scan. A novel optimized and simple configuration for the minimization of this effect is also proposed and experimentally validated.

\section{Results and discussion}
\label{sec:results}

Despite its apparent simplicity, as a direct consequence of Snell's law, the refraction of light at a flat boundary between two media produces an effect similar to spherical aberration, so that the rays coming from a point in one media will not converge again at another point after refraction. Instead, all of them will be tangent to a curve, the envelope of the refracted rays (in general, a surface in three dimensions), so called \textit{caustic}. Therefore, if we scan the refracted rays, the image point appears to move along the caustic surface \cite{southall_mirrors_1918}. This is a well known issue in geometrical optics and can be applied to understand the problems arising when a prism is rotated around a fixed axis. Indeed, if we draw all the refracted rays that would meet at a certain position on the base of the prism after refraction at the other face, the corresponding incident rays will never meet at a single point. They will generate a caustic curve which is part of an astroid with a symmetry axis along the direction of normal incidence, as depicted in Fig. \ref{fig:caustic}. So, clearly, with a fixed rotation axis for the incident beam or, equivalently, for the prism stage, it is impossible to perform angle-scanning and maintain a completely stationary interrogation spot on the sensing surface.

The rotation axis is usually depicted and, very commonly, experimentally set in the center of the prism base. As we will show below, in this situation and in many other frequent choices, the walk-off during angle scan can not be neglected in many cases. It is worth noting here that in the case of a semicylindrical prism, stationarity can be ensured over any angle scan but only for the central point. Interrogation of any other point, i.e. arraying, would also suffer from walk-off.

For any type of prism, getting rid of the walk-off during interrogation requires performing a translation every time the angle is shifted. This solution has been theoretically described by Fontana el al. \cite{fontana_maintaining_2013} and experimentally implemented by Singh et al. \cite{singh_fixed_2014}. However, this process, although it resolves the issue, it does so at a high price, because it requires very high precision positioning and can be very time-consuming, specially in situations where extensive sample arraying is carried out (in \cite{singh_fixed_2014} three different mirrors have to be precisely adjusted for every angle measured).

\begin{figure}
\begin{center}
  \includegraphics[width=\linewidth]{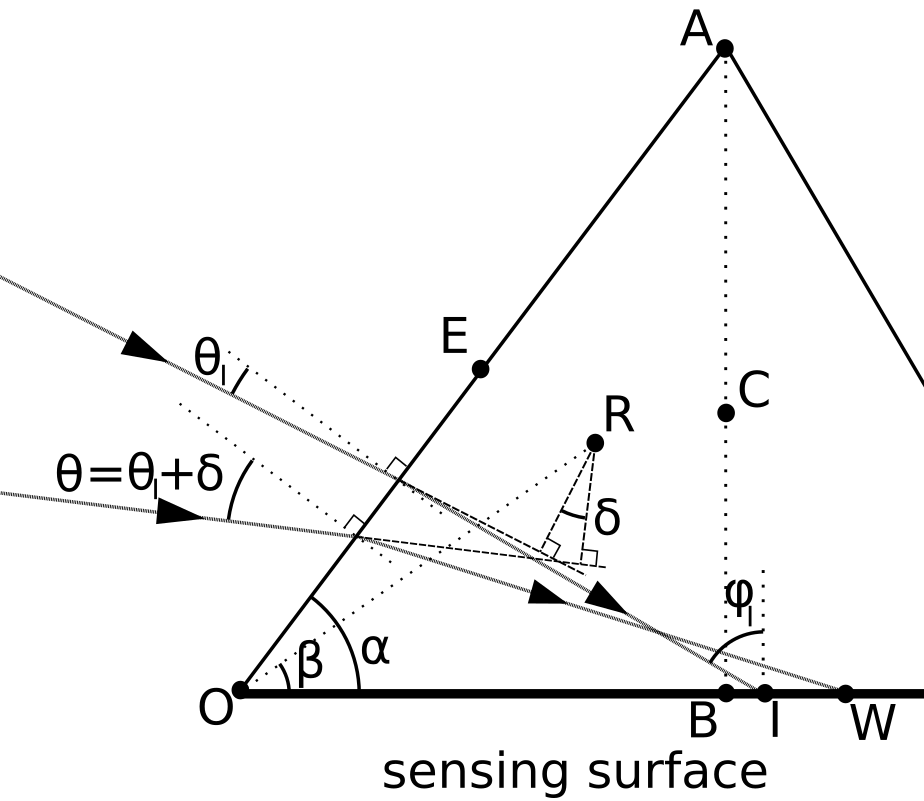}
\end{center}
\caption{Scheme of the notation employed in the tracing of two rays incident on the prism related by a $\delta$ rotation around R. The prism has an opening angle $\alpha$. The incident ray with angle $\theta_I$ hits the sensing surface at I. After the $\delta$ rotation, the ray will hit at W, giving rise to the walk-off of the beam along the sensing surface $IW$. The rotation axis R is determined by the angle $\beta$ and the distance $OR$. Some additional positions are marked and defined in the figure.}
\label{fig:trian}
\end{figure}

In any case, there exists a position of the rotation axis in which the walk-off is minimized for a given angle scan interval. This position depends on the geometry of the prism and the angle scan interval of interest. Ulrich et al.  proposed a setup with two different rotation axis positions depending on the sign of the angle of incidence with respect to the prism side normal ($\theta$ in Fig. \ref{fig:trian}) \cite{ulrich_measurement_1973}. Any angle scan going through $\theta=0$ would need a discontinous repositioning of the rotation axis giving rise to artifacts in the measurement. More recently, Qi et al. also analyzed the problem and proposed an alternative rotation axis position based on symmetry and geometric considerations \cite{qi_optimized_2012}. To the best of our knowledge, no other quantitative analysis is found in the literature. As we will see, although both optimization methods result in improvements of some of the above-mentioned typical choices, these positions do not ensure minimum walk-off for angle scan experiments. 

We perform here a complete analysis of the interrogation point walk-off with an arbitrary position of the rotation axis R for a triangular prism. Although ray-tracing software can be used to evaluate the magnitude of the deviation from the interrogation point for a certain configuration, we want to analyze it for any rotation axis, so that we can find the position of R for which that deviation is minimum throughout a whole scan. Therefore, we do a complete analytical survey of the optical path followed by the incident beam. The scheme and employed notation are explained in Fig. \ref{fig:trian} where the section of the prism corresponding to the plane of incidence is depicted. The incident beam is not restricted to point towards R so that it can be offset. We define the position of R with the distance $OR$ and the angle $\beta$ and we suppose that the axis is normal to the depicted surface. The opening angle of the beam side of the prism is $\alpha$. The desired interrogation point (point of interest) in the sensing face of the prism is I, which is at a distance $OI$ from the left corner O. This point I is exactly interrogated when the angle of incidence at the side face of the prism is $\theta_I$. The corresponding incidence angle at the sensing surface inside the prism is called $\phi_I$. If we rotate the beam around R an angle $\delta$ from this state, the incoming beam will have an incidence angle $\theta=\theta_I+\delta$  and will hit the sensing surface at W, resulting in a walk-off $IW$ (see supplementary information for an analytical expression of $IW$).

\begin{figure}
\begin{center}
  \includegraphics[width=\linewidth]{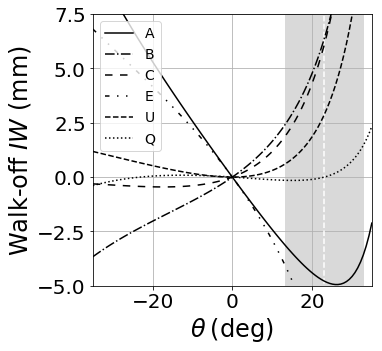}
\end{center}
\caption{Walk-offs $IW$ as a function of the incident angle $\theta$ calculated for different positions of the rotation axis R. A, B, C and E correspond to those of Fig. \ref{fig:trian}. U and Q correspond to proposals made in \cite{ulrich_measurement_1973} and \cite{qi_optimized_2012} respectively. The typical angle interval scanned for measuring the SPR dip in a water-based environment is shaded in grey. The resonance angle is highlighted with a white dashed line. }
\label{fig:ODinitial}
\end{figure}

Before continuing we want to highlight the order of magnitude of this undesired effect during angular scanning. In Fig. \ref{fig:ODinitial} we have represented the walk-off $IW$ as a function of incidence angle $\theta$ for various common positions of R (see Fig. \ref{fig:trian}) and the optimized values found in the literature in a typical setup with an equilateral glass prism (n=1.51, $OA=25$ mm) with  $\alpha=60^\circ$ supposing a gold sensing surface and $\lambda=633 $ nm. The point of interest to be interrogated is taken at $OI=12.5$ mm (center of the sensing surface). The geometry has been set up so that in all cases the walk-off at normal incidence is null, which is the most common choice even in optimized setups \cite{ulrich_measurement_1973,horn_plasmon_2010,qi_optimized_2012}. As it can be seen from the figure, when varying the incidence angle the walk-off increases. When the rotation is carried out around points A, B, C, E and U it does so in a way that the beam never hits the desired spot I again during the scan shown. In the case of positioning the rotation axis in Q as suggested by Qi eta al. \cite{qi_optimized_2012} the walk-off is more controlled. In any case, the walk-off can be an important effect of the order of magnitude of millimeters and larger. 

It is worth noting that in practice the angle interval of interest is always more or less centered at the angle of minimum reflectance (resonance angle) and wide enough to allow a good determination of the minimum position. Typically this interval could be of 20$^{\circ}$. In the case of the prism defined above, this interval would be centered at $\theta_{\mathrm{res}}\sim 23^\circ$ for sensing in a water-based medium $n\sim 1.33$ (corresponding to $\phi_{\mathrm{res}}\sim 75^\circ$). The interval is shown shaded in Figure \ref{fig:ODinitial} (the reflectivity dip due to the plasmon resonance is explicitly shown in Fig. \ref{fig:IW_op}). As it can be seen, even within this reduced interval, the walk-offs are non-negligible. From the options shown in the graph, the best choice would be the position Q with a walk-off of $\sim 2$ mm around the spot of interest I. This position is in fact calculated so that at $\theta_\mathrm{res}$ $IW=0$ (see \cite{qi_optimized_2012}). In other cases the walk-off is much larger and/or the spot of interest is not even interrogated which, in effect, results in a poorly designed experiment. 

\begin{figure}
\begin{center}
  \includegraphics[width=\linewidth]{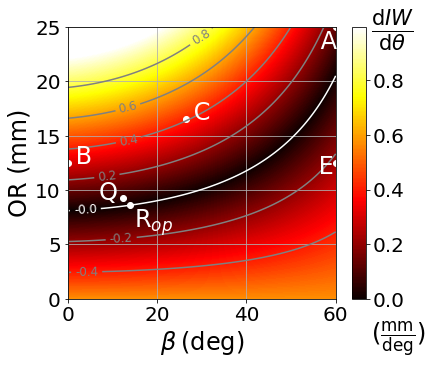}
\end{center}
\caption{Variation of the walk-off $IW$ with respect to the angle of incidence $\theta$ $\left(\frac{\mathrm{d}IW}{\mathrm{d}\theta}\right)$ for the case mentioned in the text as a function of the position of R, given by the distance OR and the angle $\beta$. The particular points shown in the graph correspond to the positions of the rotation axis according to the notation in Fig. \ref{fig:ODinitial}; R$_\mathrm{op}$ is the best position in order to hit the point of interest when the angle of incidence is interrogated at the resonance angle.}
\label{fig:cmap}
\end{figure}

\begin{figure}
\begin{center}
  \includegraphics[width=\linewidth]{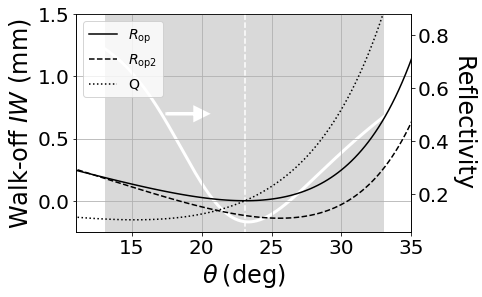}
\end{center}
\caption{Graph showing the walk-off $IW$ as a function of the angle of incidence $\theta$ in the interval of interest (shaded region) around the angle of resonance (dashed white line) for the case defined in the text. Continuous line: rotation around our optimized position R$_\mathrm{op}$. Dashed line: rotation around R$_\mathrm{op2}$ so that the walk-off is symmetrical with respect to the point I. Dotted line: rotation around the optimized position proposed in \cite{qi_optimized_2012}. The curve of reflectivity for a SPR prism with a gold sensing layer ($\epsilon_\mathrm{Au}=-9.81+1.96\mathrm{i}$) in a water-based environment ($n_{\mathrm{H_2O}}=1.33)$ is also shown as a white line to prove that the shaded region is effectively where the SPR reflectivity dip happens.}
\label{fig:IW_op}
\end{figure}

We will base our analysis in the minimization of the walk-off $IW$ in the angle interval of interest removing the common constraint of interrogating the desired spot at $\theta=0$ since, in many cases, this angle is not of interest. Moreover, we impose that $\frac{\mathrm{d}IW}{\mathrm{d}\theta}$ is null at $\theta_{\mathrm{res}}$ to ensure that the walk-off $IW$ is at a stationary point at the resonance angle. This derivative is represented as a heatmap with respect to the position R defined by parameters $OR$ and $\beta$ in Fig. \ref{fig:cmap}. As expected, there are positions of R providing more stable $IW$s than others. Concretely, the position of the minimum, corresponding to the contourline of null value, guarantees that as we scan I at $\theta_\mathrm{res}$, $IW$ will undergo its minimum rate of change.

We can take advantage of Fig. \ref{fig:cmap} to understand the criticality of the rotation axis position R in the proposed configuration. At first order, positioning R $3$ mm away from the positions defined by the contourline of null value (shown in white in Fig. \ref{fig:cmap}) can result in a walk-off of 0.2 mm per degree. Taking into account that the scan width can be of $20^\circ$, the effect can be of large importance. In the figure we have also represented the points corresponding to setting R at the particular points defined in Fig. \ref{fig:trian} to show that the best option Q is the closest to the mentioned contourline. In order to minimize $IW$, we need to determine the exact best position for R along the white contourline. We use the fact that our point of interest is I and we make  $\theta_I=\theta_{\mathrm{res}}$ so that I is exactly interrogated at the resonance angle in the middle of the angle scan. This is achieved for $OR=8.6$ mm and $\beta=13.9^\circ$ marked in the figure as R$_{op}$.

We now calculate the walk-off $IW$ as a function of $\theta$ for this rotation axis position R$_{op}$. It is represented as the solid line in Fig. \ref{fig:IW_op}, where only the mentioned angle interval of interest around the resonance angle is shown. As we can see the minimum is achieved at the resonance angle, where $IW$ is null. Moreover, it is more constrained than for the positions of R shown in Fig. \ref{fig:ODinitial} in the same angle interval (shaded in grey). For reference, in Fig. \ref{fig:IW_op} the value for setting R at Q is also shown as a dotted line. We can define now the total walk-off as the length scanned during an angular scan. In the case of positioning R at Q, for the interval of this example, the total walk-off is 1.7 mm. Performing the scan around R$_{op}$ results in a total walk-off of 0.7 mm, 2.4 times lower than the previous result.

It is worth noticing that for R$_{op}$, $IW$ is always on the positive side and it increases faster at larger angles. This means that during the angle scan, only points to the left of I (see Fig. \ref{fig:trian}) are being interrogated. However, slightly moving the rotation axis position from R$_{op}$, a more symmetric and smaller walk-off $IW$ can be achieved. This is shown as the dashed line in Fig. \ref{fig:IW_op} where the position of R has been moved to $OR=8.5$ mm and $\beta=14.5^\circ$ so that $IW$ is equal at both scanning angle boundaries and the same distance is scanned to both sides of I. This point is called R$_\mathrm{op2}$ and gives rise to a more symmetric and further reduced total walk-off of 0.4 mm. Again, we would like to underline here the criticality of the rotation axis position R, as it is shown by the significant differences in Fig. \ref{fig:IW_op} for three positions less than 1 mm apart.

The maximum walk-off is of importance as it sets a limit in the arraying density to be achieved in angle scan experiments. A reduction of the walk-off of a factor of 4.3, such as the one of the example, results in the possibility of increasing the spot density in 2D sample arrays by a factor of 18. Interestingly enough, knowing the total walk-off a minimum spot can be defined and taken into account when preparing the sample.

In order to experimentally verify the results obtained in the previous analysis and with the aim of proposing a setup for implementation of the rotation axis adjustment, we have measured the walk-off $IW$ employing the arrangement shown in Fig. 6. The prism is an isosceles right angle BK7 glass prism ($\alpha=45^\circ$, $OA=20$ mm) with two polished faces and one fine grounded for light scattering. Its refractive index is 1.5151 at the wavelength of illumination (diode laser at 635 nm). The prism is fixed on the laboratory frame of reference and the scattered light from non-polished face is observed via a CCD camera. A motorized rotatory stage is mounted on a XY translation stage and used to rotate an optical rail where a diode laser and focusing lens are mounted. The beam, initially aligned along the rail, can be off-set with a translation stage. This setup allows arbitrary positioning of the rotation axis R while the beam can be off-set to interrogate any point on the prism base at any given angle.

For SPR sensing in a water-based medium where the mentioned prism would be coated with a gold sensing thin film, the resulting resonance angle is $\theta_{\mathrm{res}}=48.3^\circ$ (corresponding to $\phi_{\mathrm{res}}=75.1^\circ$). Using the described setup, we measured the walk-off $IW$ when interrogating the center of the sensing surface B in the angle interval of interest $\left[38^\circ,58^\circ\right]$ while rotating around three different R positions (the same as in Fig. \ref{fig:IW_op}). In Fig. \ref{fig:IW_Esp} we show the experimental data together with the theoretical curves. As it can be seen, the agreement is good. As expected, the minimum total walk-off is obtained for the position  R$_\mathrm{op2}$ proposed in the present paper.

It is worth noting the order of magnitude of the walk-off in this experiment. Even in the optimized setup the total walk-off is c. 0.9 mm, which can easily increase to several millimeters if R is positioned in non-optimum positions (for example to more than 7 mm for the circle data points in Fig. \ref{fig:IW_Esp}). One of the reasons of this large dependence of the walk-off on the positions of R is due to the fact that the interrogation is being carried out far from normal incidence on the prism side. Optimized prisms can be designed with opening angles $\alpha$ and refractive indices appropriate for a given sensing medium and interrogation wavelength so that $\theta_\mathrm{res}$ is close to zero, since in the paraxial approximation (or even for small angle intervals far from  normal incidence) the caustic curve of incoming rays almost collapses to a point. Although this can reduce the walk-off to a great extent, the effect will always be present and has to be taken into account when designing the experiment. On a second note but of high importance, the widespread and sometimes indiscriminate use of off-the-shelf commercial prisms with $\alpha=45^\circ$ or $60^\circ$ results in experiments carried out in poor or non-controlled conditions.

\begin{figure}
\begin{center}
  \includegraphics[width=\linewidth]{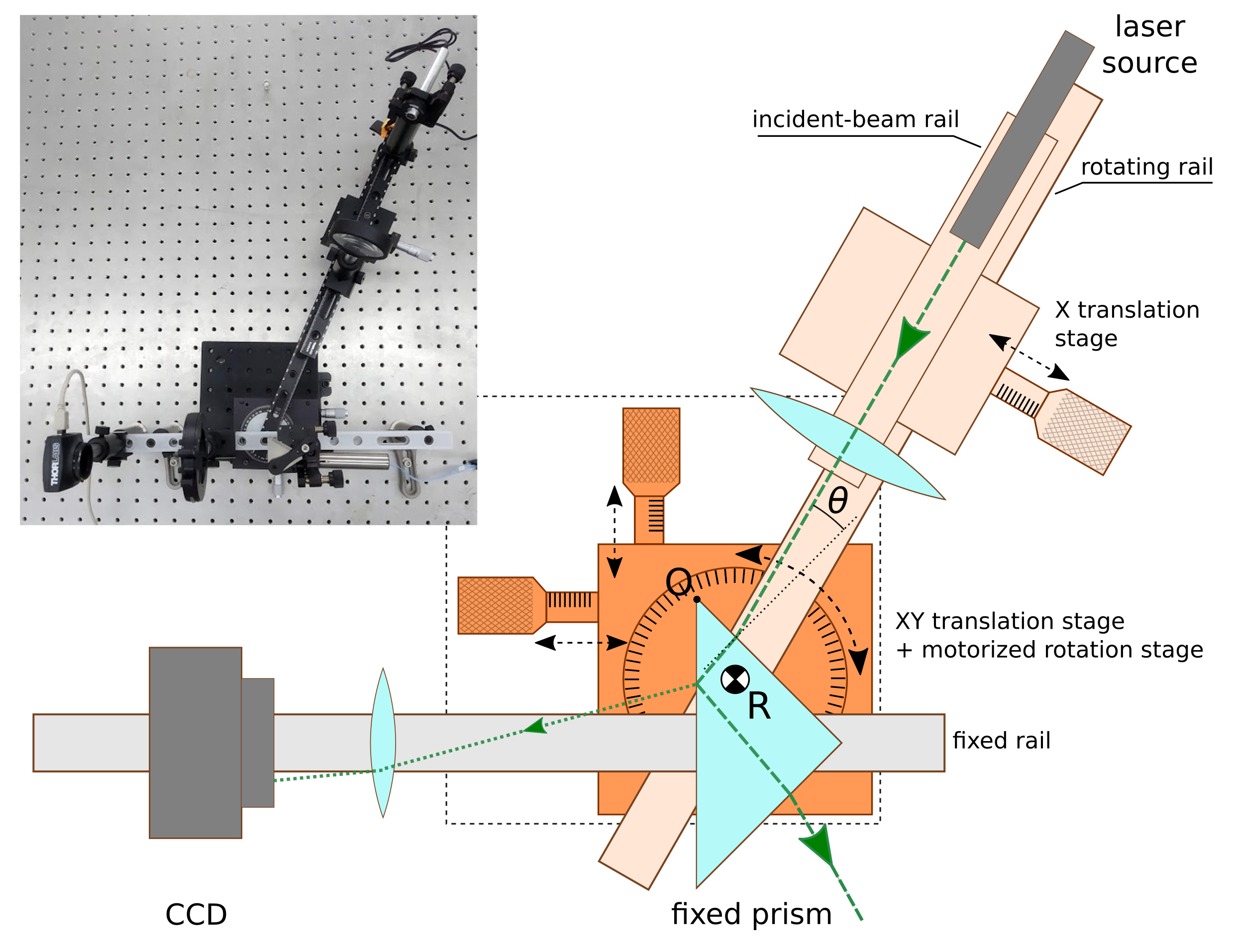}
\end{center}
\caption{Scheme of the setup used to measure the walk-off $IW$ as a function of the angle of incidence $\theta$. The position of the rotation axis R with respect to the laboratory fixed prism can be set with a XY translation stage on which a motorized rotation stage is mounted. A laser diode is mounted on the rotating rail and can be off-set via a linear translation stage. Inset: Real experimental setup on the optical table.}
\label{fig:setup}
\end{figure}

\begin{figure}
\begin{center}
  \includegraphics[width=\linewidth]{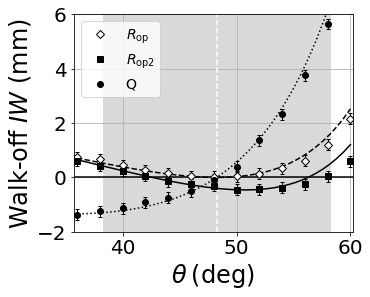}
\end{center}
\caption{Theoretical curves (lines) and experimental data (points) for the walk-off $IW$ vs. $\theta$ in the interval of interest (shaded region) around the angle of resonance (dashed white line) for the experimental configuration described in the text: dashed line and empty rhombi are obtained for the rotation around our optimized position R$_\mathrm{op}$; for the continuous line and squares the rotation axis is shifted from R$_\mathrm{op}$ to R$_\mathrm{op2}$ in order to get a symmetrical walk-off around the point of interest; and dotted line and circles correspond to the rotation around Q.}
\label{fig:IW_Esp}
\end{figure}

Finally, we would like to point out that the results here presented are not restricted to Kretschmann-like SPR setups and can potentially be applied to other experimental techniques based on prism-coupling and total internal reflection.
\section{Conclusions}
Although the effects of the prism rotation are usually neglected in SPR experiments, we have shown that the walk-off of the spot on the sensing surface in configurations such as Kretschmann's should be taken into account, especially when the sample is not homogeneous (e.g., defects in thin films or sample arrays for multiplexing). First, we have obtained the value of the walk-off as a function of the angle of incidence of light for any position of the rotation axis. We have concluded that it strongly depends on the position of the rotation axis and the characteristics of the prism (opening angle and refractive index). However, we have also shown that there exists an optimal position for the rotation axis such that the value of the walk-off is minimized at the angle of SPR. By requiring a symmetrical walk-off around the point of interest, we proposed an even better position of the rotation axis which results in a further reduction of the total walk-off for the angular interval of interest. In order to show the ease of implementation of the proposed configuration we experimentally verified the setup and results. Lastly, it is noteworthy that these results and conclusions are not restricted to SPR systems, but may be applied to any technique using prism-coupling with varying angles of incidence.

\bibliography{SPRbib}

\end{document}